\documentclass[twocolumn]{aastex631}
\usepackage{subfigure}

\usepackage{newtxtext,newtxmath}
\usepackage[T1]{fontenc}
\DeclareRobustCommand{\VAN}[3]{#2}
\let\VANthebibliography\thebibliography
\def\thebibliography{\DeclareRobustCommand{\VAN}[3]{##3}\VANthebibliography}
\usepackage{graphicx}
\usepackage{amsmath}
\usepackage{natbib}
\usepackage{multirow}
\usepackage{subfigure}
\usepackage{hyperref}
\usepackage{textcomp}

\newcommand{\Rmnum}[1]{\expandafter@slowromancap\romannumeral #1@}
\makeatother
\begin{document}
\title{Observations of the X-ray Millihertz Quasiperiodic Oscillations in Hercules X-1}

\author{Wen Yang}
\affiliation{Department of Astronomy, School of Physics and Technology, Wuhan University, Wuhan 430072, China}
\author{Wei Wang}
\altaffiliation{Email address: wangwei2017@whu.edu.cn}
\affiliation{Department of Astronomy, School of Physics and Technology, Wuhan University, Wuhan 430072, China}
\begin{abstract}
With a systematic timing investigation of the persistent X-ray binary pulsar Her X-1 based on a large number of Insight-HXMT observations between 2017 to 2019, we confirm the presence of X-ray millihertz quasi-periodic oscillations (mHz QPOs) at $\sim 0.01$ Hz. By applying wavelet analysis in our data analysis procedures, we firstly identified $\sim 0.005-0.009$ Hz QPOs coexisting with the $\sim 0.01$ Hz QPOs. Wavelet analysis suggests that these QPO features show transient behaviors, frequencies of mHz QPOs evolved in short time scales. There exists a positive relation between QPO centroid frequency (from $\sim 0.005-0.009$ Hz) and the X-ray luminosity, while the 10 mHz QPO frequencies keep nearly constant for different luminosities, which suggests different physical mechanisms for two types of mHz QPOs. The 10 mHz QPOs in both X-ray and UV bands would have the same origin related to the beat frequency where the Alfv$\acute{e}$n radius is close to the corotation radius, and the 5 mHz QPOs may originate from magnetic disk precession.
\end{abstract}
\section{Introduction} \label{sec:intro}
Hercules X-1 (Her X-1) is a bright persistent X-ray binary pulsar system, which was discovered in 1972 using Uhuru \citep{tananbaum1972discovery}. This intermediate-mass system hosts a highly magnetized 1.4$M_{\odot}$ neutron star with a 1.24 s spin period in a 1.7-day orbit with its approximately $2M_{\odot}$ optical companion, HZ Herculis \citep{reynolds1997new}. Similar to X-ray binaries with high-mass companions (HMXBs), this system shows pulsations, and its optical luminosity is primarily driven by the companion's emission \citep{vrtilek2001multiwavelength}. The orbit is nearly circular with a high inclination angle of $85^\circ$ causing a six-hour long eclipse \citep{staubert2009updating}. 
\begin{table*}
\centering
\caption{Details on Insight-HXMT observations of Her X-1 from 2017 to 2019. The observation ID(obs ID) are the last two digits of P01013080, and the dash '-' represents all the observations between two obs IDs.}
\label{table1}
\resizebox{0.96\textwidth}{!}{ 
\begin{tabular}{cccccccc}
\hline
\begin{tabular}[c]{@{}l@{}}Observation\\ (year/month)\end{tabular} & obs ID & 35-day cycle\footnote{The counting of 35-day cycle follows the convention proposed by \citet{staubert1983hercules}}
& \begin{tabular}[c]{@{}l@{}}Start of obs\\ {[}MJD{]}\end{tabular} & \begin{tabular}[c]{@{}l@{}}End of obs\\ {[}MJD{]}\end{tabular} & \begin{tabular}[c]{@{}l@{}}Center of obs\\ {[}MJD{]}\end{tabular} & \begin{tabular}[c]{@{}l@{}}Exposure\\ {[}ksec{]}\end{tabular} & \begin{tabular}[c]{@{}l@{}}35-day\\ phase \footnote{35-day phase is calculated based on the center of observations with the reference time for the zero phase at MJD 53753.1988, corresponding to the peaks of the 35-day cycle \citep{leahy2020swift}.}
\end{tabular} \\ \hline
2017 July & 01 & 477 & 57935.14 & 57936.95 & 57936.04 & 40 & 0.02 \\
2017 August & 02-03 & 478 & 57970.29 & 57971.95 & 57971.12 & 23 & 0.03 \\
2018 January & 04-05 & 483 & 58138.92 & 58141.52 & 58140.22 & 55 & 0.88 \\
2018 February & 06 & 484 & 58175.44 & 58175.81 & 58175.62 & 10 & 0.89 \\
2018 May & 07-12 & 486 & 58243.01 & 58249.04 & 58246.02 & 39 & 0.92 \\
2018 June & 13-16 & 487 & 58276.24 & 58279.64 & 58277.94 & 55 & 0.83 \\
2018 August & 17-21 & 489 & 58345.05 & 58349.68 & 58347.36 & 16& 0.83 \\
2019 February & 22-25 & 494 & 58517.21 & 58542.68 & 58529.95 & 46 & 0.06 \\
2019 May & 26 & 495 & 58556.20 & 58556.72 & 58556.46 & 9 & 0.83 \\ \hline
\end{tabular}
}
\end{table*}
\par
The system Her X-1 has been studied in X-rays, optical, and ultraviolet bands \citep{leahy2020astrosat,leahy2003modelling,jimenez2002high}. The 35-day cycle is a prominent feature of Her X-1 \citep{leahy2011light}, arising from the precession of the accretion disk around the neutron star. This precession periodically obscures the neutron star, resulting in substantial X-ray variability \citep{petterson1975hercules}. The superorbital modulation manifests in four distinct phases: an 11-day main-on state characterized by a rapid increase in brightness, variable peak durations, and a gradual decline; an 8-day short-on state; and two intervening 8-day off states, during which the X-ray flux diminishes to nearly zero \citep{scott1999rossi}. 
Her X-1 is also remarkable because of the first report of the cyclotron resonant scattering feature in its X-ray spectrum near 40 keV in 1977 with balloon observations \citep{trumper1978evidence}, implying a magnetic field strength $\sim 2.9\times 10^{12}\,\,G$ \citep{xiao2019constant}.
\par
Another type of variability, known as millihertz quasi-periodic oscillations (mHz QPOs), is observed in HMXBs. Two primary models associated with the inner accretion disk are used to explain mHz QPOs in X-ray pulsars (XRPs): the Beat-Frequency Model (BFM; \citealt{alpar1985gx5}) and the Keplerian Frequency Model (KFM; \citealt{van1987intensity}). QPOs have been detected in several high-mass X-ray binary (HMXB) pulsars, concentrated in the low-frequency range from $\sim$ 10 mHz to $\sim$ 1 Hz, including 4U 0115+63 (10 mHz, 22 mHz, 41 mHz and 62 mHz; \citealt{ding2021qpos}), KS 1947+300 (20 mHz; \citealt{james2010discovery}), IGR J19294+1816 (30 mHz; \citealt{raman2021astrosat}), Cen X-3 (40 mHZ; \citealt{liu2022detection}), SAX J2103.5+4545 (44 mHz; \citealt{inam2004discovery}), 1A 0535+262 (50 mHz; \citealt{finger1996quasi}), V 0332+53 (51 mHz; \citealt{takeshima1994discovery}), XTE J1858+034 (110 mHz; \citealt{paul1998quasi}), EXO 2030+375 (200 mHz; \citealt{angelini1989discovery}), XTE J0111.2-7317 (1.27 Hz; \citealt{kaur2007quasi,devasia2011rxte}). However, QPOs in HMXB pulsars are rare and transient phenomena. Even in systems where QPOs have been observed, they are not consistently detected during every X-ray outburst \citep{paul2011transient}.
\par
Some mHz QPOs of multi-wavelength bands in X-ray pulsars are also reported, which would be more helpful to understand physical mechanism for these mHz QPOs. In transient X-ray pulsar $4U 1626-67$, 48 mHz QPOs were found in both X-ray and optical bands, and the optical QPO is likely caused by reprocessing off the interface of the strongly irradiated companion star of an X-ray QPO \citep{chakrabarty1998high}. Her X-1 was found to show mHz QPOs in its ultraviolet (UV) continuum at frequencies of $8\pm 2$ and $43\pm 2$ mHz using Hubble Space Telescope, probably emitted on the heated atmosphere of HZ Her \citep{boroson2000discovery}. A series of low-frequency optical quasi-periodic oscillations (QPOs) with a central frequency near 35 mHz have also been observed with Keck Observatory \citep{o2001keck}. Evidence for some excess power at 10 mHz in Her X-1 is possibly interpreted as X-ray QPOs based on RXTE observations in 1996 \citep{moon2001discovery} and 1998 \citep{o2001keck}.
\par
At present, there is still the lack of the solid evidence for X-ray mHz QPOs in Her X-1. In this paper, the light curve of Her X-1 is utilized to search for the millihertz QPOs with Insight-HXMT data from 2017 July to 2019 May. Section \ref{OBSERVATIONS} describes the observations and data reduction methods. Section \ref{DATA ANALYSIS AND RESULTS} provides the details of data analysis, and the QPO results based on the power density spectrum and wavelet methods. We discuss the physical origins of the observed QPOs in Section \ref{DISCUSSION}.

\section{OBSERVATIONS}
\label{OBSERVATIONS}
The Hard X-ray Modulation Telescope(Insight-HXMT) is the first X-ray astronomical satellite in China \citep{zhang2014introduction}, launched on 2017 June 15. Insight-HXMT consists of three main instruments: the High Energy X-ray telescope (HE) operating in 20--250 keV with the geometrical area of 5100 cm$^2$ \citep{liu2020high}, 
the Medium Energy X-ray telescope (ME) operating in 5--30 keV with a geometrical detection area of 952 cm$^{2}$ \citep{cao2020medium} and the Low Energy X-ray telescope (LE) covering the energy range 1--15 keV with a geometrical detection area of 384 cm$^2$ \citep{chen2020low}.
\par
From 2017 July to 2019 May, Insight-HXMT was triggered with 9 observations, focusing on the main-on states of Her X-1 (from 35d cycle No.477 \footnote{The counting of 35d cycles follows the convention used by \citet{staubert1983hercules}: (O–C) = 0 for cycle No. 31 with turn-on near JD 2442410.}) with total exposure time of $\sim 300 ~\rm ks$. The zero phase is located at the peaks of the 35-day cycle and the reference time is 53753.1988 derived by \citet{leahy2020swift}. Details of those observations are summarized in Table\ref{table1}. 
\par
The Insight-HXMT Data Analysis Software (HXMTDAS) v2.04 is used to analyze data (more details on the analysis were introduced in previous publications, e.g., \citealt{WANG20211,zhu2024energy}). To take advantage of the best-screened event file to generate the high-level products including the energy spectra, response file, light curves and background files, we use tasks $he/me/lepical$ to remove spike events caused by electronic systems and $he/me/legtigen$ be utilized to select good time interval (GTI) when the pointing offset angle $< 0.04^\circ$; the pointing direction above earth $> 10^\circ$; the geomagnetic cut-off rigidity $>8$ GeV and the South Atlantic Anomaly (SAA) did not occur within 300 seconds. Tasks helcgen, melcgen, and lelcgen are used to extract X-ray lightcurves with $\sim 0.008(1/128)$ sec time bins.
\begin{figure*}
    \centering
    \subfigure{\includegraphics[width=.48\textwidth,height=6cm, keepaspectratio]{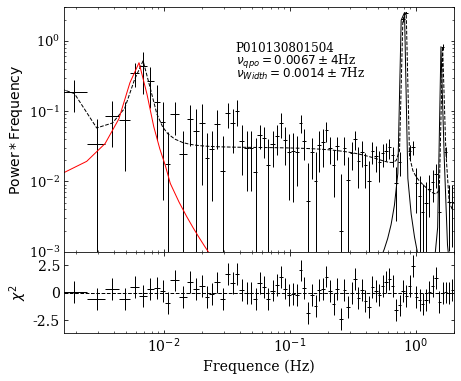}
    }
    \subfigure{\includegraphics[width=.48\textwidth,height=6cm, keepaspectratio]{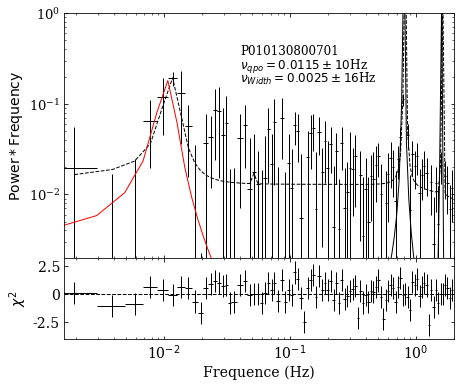}
    }
    \caption{The power density spectra for the two representative observations using the Insight-HXMT/ME data (10-30 keV). The solid lines show the best fit with a multi-Lorentzian function (dotted lines). Observation ID, the frequency and width of QPO are shown for each panel.}
    \label{fig:1}
\end{figure*}
\section{DATA ANALYSIS AND RESULTS}
\label{DATA ANALYSIS AND RESULTS}
\subsection{Power density spectrum}
We first carried out background subtraction processes on the extracted light curves obtained from each payload and exposure. Powspec from HEASOFT was employed to calculate the power density spectra (PDS) for the light curves in each observation, using a time interval of 1024 seconds and a corresponding time resolution of 1/128 seconds. We generated PDSs in the energy bands of 1-10 keV for LE data, 10-30 keV for ME data. The data with energy above 30 keV were excluded due to the photon count rate being less than 10 counts/s. The PDSs were averaged and normalized to ensure their integral equals the square of the RMS fractional variability. Subsequently, we identified observations with clear QPO signals. 
\par
The PDSs can be characterized by a model generally consisting of a broad-band noise, a spin frequency peak at $\sim 0.8$ Hz, its harmonic at $\sim 1.6$ Hz, and a signature of QPO bump at $\sim 0.01$ Hz. To infer the fundamental parameters of the QPO, the PDSs were fitted with multiple Lorentz functions to describe broadband noise, QPO, and two spin components. The fractional root mean square (rms) of the QPO can be summarized as 
\begin{equation}
rms_{QPO}=\sqrt{R}\times \frac{S+B}{S},
\end{equation}
where S represents the source count rate, R represents the normalization of the QPO's Lorentzian component, and B represents the background count rate \citep{bu2015correlations}.

The examples based on the ME observations of P0101308015 and P0101308007 are presented in Figure \ref{fig:1} respectively. The average QPO centroid frequency was determined to be $0.0115\pm 0.0010$ Hz for P01030800701, and $0.0067\pm 0.0004$ Hz for P01030801504. The quality factor $Q=\frac{v}{\bigtriangleup v}$ (where $v$ represents the frequency of the QPO and $\bigtriangleup v$ represents the full width at half maximum; FWHM) is calculated to be $\sim 4$  and the white noise subtracted RMS value of the QPO in the 10-30 keV is determined to be $\sim 3\%$ for the observations. We also test the fitting when excluding the Lorentz function near the 10 mHz QPO feature for the ObsID P010130800701. The value of $\chi^2$ changed from 84 (80 d.o.f., with the Lorentz function centered at $\sim$ 10 mHz included in the model) to 112 (83 d.o.f., without the 10 mHz Lorentz function).

Due to the short GTIs (less than 800 seconds) in the LE band (1-10 keV) for the observations, the statistical significance of the 10 mHz signal is low. Consequently, we employed wavelet analysis to systematically investigate the millihertz QPO, allowing for the detection of the signal within short time intervals by providing both time and frequency localization, which is particularly advantageous for capturing transient and non-stationary signals that may otherwise be missed by traditional Fourier analysis.
\subsection{Wavelet analysis}
Wavelet analysis is similar in essence to the Fourier transform as a signal processing technique used to decompose and analyze the frequency components of a time series. However, there are important differences between the two approaches. The Fourier transform assumes the signal is stationary and all-time domain information is lost. On the other hand, the basis functions of wavelet transform are finite in duration and can be scaled and translated, providing a more localized and flexible representation of the signal and therefore can be used to identify when specific frequencies appear in the time series. \citet{torrence1998practical} and the \citet{addison2017illustrated} introduced comprehensive review of wavelet transforms with elaborated step-by-step guide and examples. Consequently, wavelet analysis should be ideal for studying X-ray light curves to analyze the time-varying frequency content. For this work, we used the pycwt \footnote{\href{https://github.com/regeirk/pycwt}{https://github.com/regeirk/pycwt}} package to compute the continuous wavelet transform (CWT). 
\par
The continuous wavelet transform $w_n\left( s \right)$ of a time series $x_{n'}$ is defined as the convolution of the time series with the complex conjugate of the wavelet basis vector:
\begin{equation}
w_n\left( s \right) =\underset{n'=0}{\overset{N-1}{\varSigma}}x_{n^{'}}\varPsi ^*\left[ \frac{\left( n^{'}-n \right) \delta t}{s} \right] .
\end{equation}
Here the * means the complex conjugate, the parameters s and n represent the scaling and shifting of the wavelet basis vector $\varPsi$. The scaling parameter s in the wavelet transform is similar to the scaling factor in the Fourier transform, as it represents each frequency component. The shift parameter n can be considered as the time, which is not present within the Fourier transform.
\par
To calculate the wavelet power spectrum of a time series, it is necessary to define a wavelet basis vector(mother wavelet). A commonly used wavelet basis Morlet in our case is chosen, and has the following form:
\begin{equation}
\varPsi _0\left( \eta \right) =\pi ^{-\frac{1}{4}}e^{i\omega_0\eta}e^{-\frac{\eta ^2}{2}} ,
\end{equation}
where $\eta$ is a time parameter, and the Morlet wavelet at a central frequency of $\omega_0$. $\varPsi _0$ means the $\varPsi$ has not been normalized. To ensure that the wavelet transforms at each scale s are directly comparable to each other and to the transforms of other time series, the wavelet function at each scale s is normalized to have unit energy:
\begin{equation}
\overset{\land}{\varPsi}\left( s\omega_k \right) =\left( \frac{2 \pi s}{\delta t} \right) ^{\frac{1}{2}}\overset{\land}{\varPsi _0}\left( s\omega_k \right) ,
\end{equation}
where the properties of $\overset{\land}{\varPsi}_0\left( s\omega_k \right)$ can be seen in Table 1 of \citet{torrence1998practical}, $\omega_k$ is the angular frequency (see Equation 5 of \citet{torrence1998practical} for the definition). Using these normalizations, at each scale s satisfy:
\begin{equation}
\underset{k=0}{\overset{N-1}{\varSigma}}|\overset{\land}{\varPsi}\left( s\omega_k \right) |^2=N .
\end{equation}
Finally, by the convolution theorem, the wavelet transform is the inverse Fourier transform of the product:

\begin{equation}
W_n\left( s \right)  =\underset{k=0}{\overset{N-1}{\varSigma}}\overset{\land}{x}_k\overset{}{\overset{\land}{\varPsi}^*\left( s\omega_k \right) e^{i\omega_kn\delta t}} ,
\end{equation}
where the * means the complex conjugate, $\overset{\land}{x}_k$ is a discrete Fourier transform of time series ($x_{n^{'}}$).  Similar to how the PSD of a time series is determined by the modulus square of the Fourier transform. The wavelet power can be defined as the modulus square of the wavelet transform $|W_n\left( s \right) |^2$ \citep{torrence1998practical}. Therefore, the wavelet power spectrum can essentially be interpreted as the modulus squared of the inner product between the time series and the wavelet basis vector for each time-frequency pairing, a diffuse two-dimensional time-frequency graph is eventually displayed as the time index and the wavelet scale change. A more detailed introduction and application of wavelet analysis in X-ray light curve can be found in \citet{ghosh2023applying} and \citet{chen2022wavelet}. 
\par
An important aspect of the wavelet power spectrum is to determine the significance level of QPO peaks in the wavelet spectrum, which can be determined by comparing the peaks in the wavelet power spectrum against the background spectrum. The method based on \citet{torrence1998practical} is outlined to estimate the red-noise background spectrum by the lag-1 autoregressive [AR(1)]. Background spectrum is calculated by multiplying the red-noise spectrum with the value of chi-square distribution, which is determined based on the given significance level (e.g., 0.95) and degrees of freedom. Finally, if a power in the wavelet power spectrum exceeds the 95$\%$ confidence level spectrum (i.e., is significant at the 5$\%$ level) and is deemed significant relative to the background spectrum. Another important aspect of the wavelet power spectrum is to understand the role of the Cone of Influence (COI), which highlights the region where edge effects are significant. 
\begin{figure}
    \centering
    \includegraphics[width=.5\textwidth]{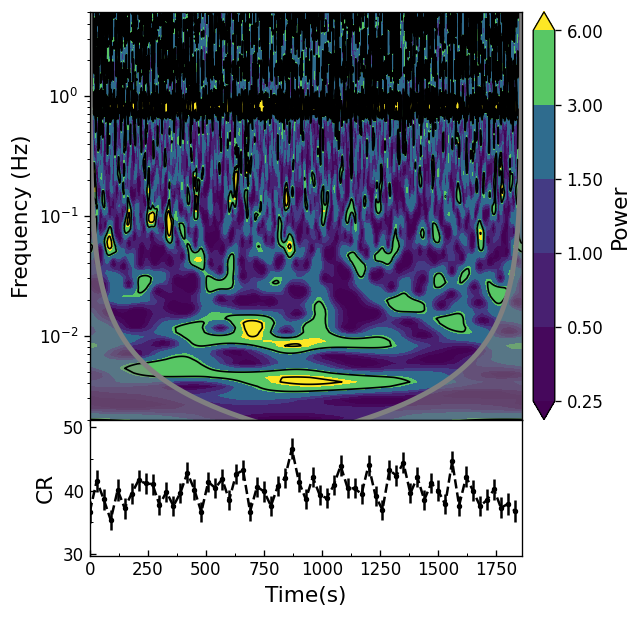}
    \caption{The wavelet result of the first GTI of Observation ID P01013080207 in ME(10-30 keV). The local wavelet spectra are presented in the top panels, with the corresponding count rates(CR) of every 30 s over the same time intervals shown in the bottom panels. Regions with greater than 95$\%$ confidence level are circled with black lines in the contour plot, and the cone of influence area is marked with gray hashed lines. Color bars of the contour plot is presented on the right side, and the value scale represents the local wavelet power. The QPO signals are found at both $\sim 5$ and 10 mHz, in addition, the 10 mHz QPO frequency possibly shows the variation in a short time scale less than 200 s.  
    }
    \label{fig:2}
\end{figure}
\begin{figure}
    \centering
    \includegraphics[width=.5\textwidth]{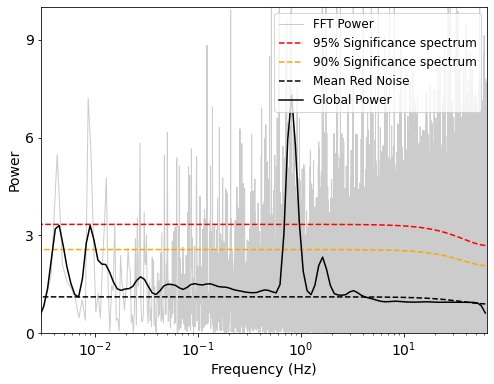}
    \caption{The global wavelet power spectrum for Her X-1 is shown for Observation ID P01013080207 during the first GTI. The solid black line is the power spectrum of the signal, which is compared to the power spectra of red noise random processes (broken lines). The detections of $\sim$ 5 mHz and $\sim$ 10 mHz QPO with the 1.86 ks observations reach the expected levels of red noise at the 95$\%$ significance levels relative to red noise with $\alpha$=0.054.
    }
    \label{fig:3}
\end{figure}
\begin{figure*}
    \centering
    \subfigure{
        \includegraphics[width=.48\textwidth, keepaspectratio]{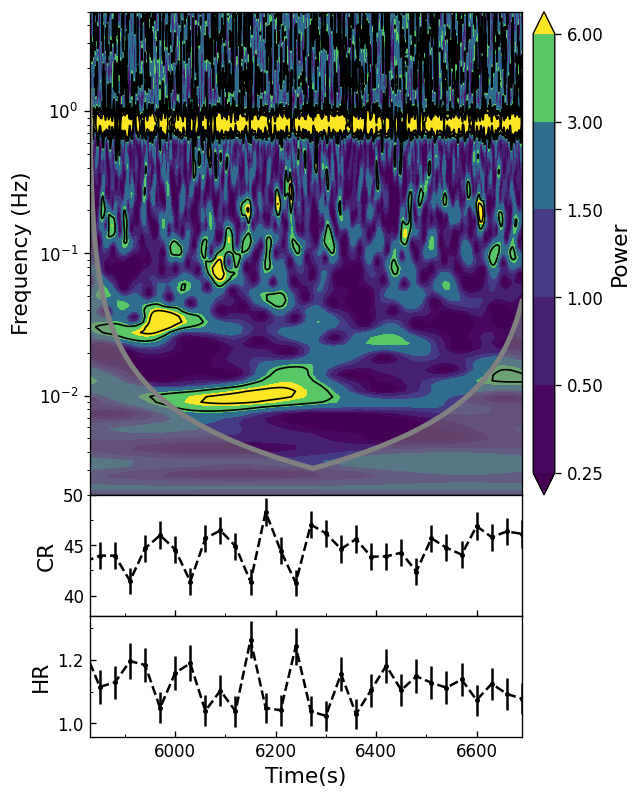}
    }
    \hfill 
    \subfigure{
        \includegraphics[width=.48\textwidth, keepaspectratio]{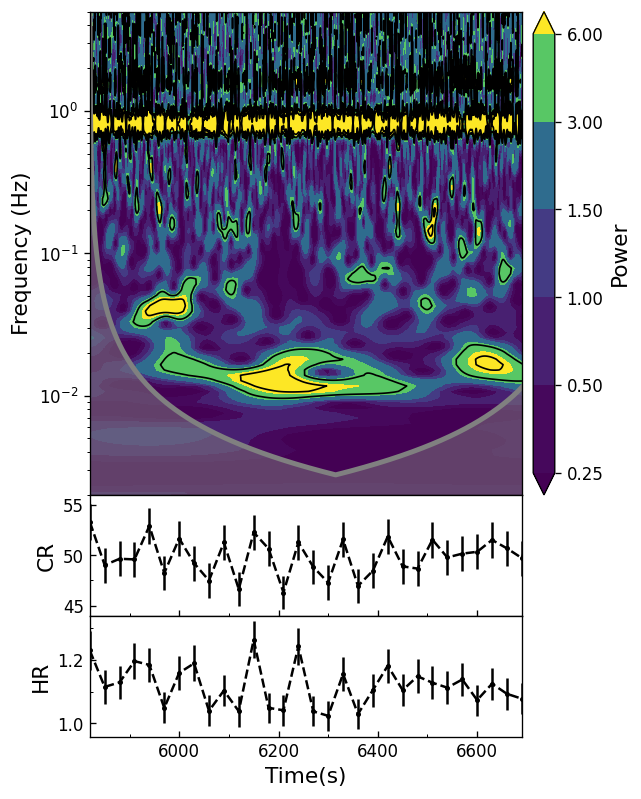}
    }
    \caption{The wavelet results, count rates (CR) and hardness ratio (10-30 keV/1-10 keV) with 30-s bins for the Obs ID P010130800507 in two energy bands: 1-10 keV (LE, left) and 10-30 keV (ME, right), respectively. The mHz QPO at LE bands has the constant frequency near 10 mHz during the observation, while the QPO frequency in the ME band shows the strong evolution from $\sim 15-18$ mHz in the early stage to the lower frequency near 10 mHz, finally becoming two frequencies at $\sim 10$ mHz and 20 mHz. For both LE and ME bands, there also exists an oscillation signal near 40 mHz lasting about 100 s in the early stage. The hashed area represents the COI. }
    \label{fig:4}
\end{figure*}

\par
Since large time intervals are presented in nearly all light curves, we selected smooth and long-duration time series for wavelet analysis to improve accuracy. Examples of the most interesting maps have been presented in Figure \ref{fig:2}. Results show a prominent QPO peak at $\sim 10$ mHz, which has also been observed in the ultraviolet (UV; \citealt{boroson2000discovery}). In addition, a $\sim 5$ mHz QPO also appeared in the wavelet power spectrum. The global wavelet power spectrum (by summing up the wavelet power spectra at all times) is presented in Figure \ref{fig:3}, which is also compared to the power spectrum of red-noise random processes. We can see that the detection of the transient QPOs in the 1.86-ks long light curve reaches the expected levels of red noise at the 95$\%$ significance levels relative to red noise with $\alpha$=0.054. To quantify the QPOs features globally, the central frequency of the QPO, full width in half maximum (FWHM) are calculated and the quality factor(Q-factor) and R-factor are estimated accordingly.

\par
Q factor has been calculated from the ratio of the central frequency to the FWHM:
\begin{equation}
Q=\frac{Centroid\,\,frequency}{FWHM}.
\end{equation}
R factor is the power of the global wavelet spectrum relative to the global 95 confidence spectrum:
\begin{equation}
R=\frac{Global\,\,signal\,\,peak}{Global\,\,95\% confidence\,\,spectrum} . \end{equation}
 For detecting the significant signals for further studies, we select the QPOs with $R$ factor greater than 0.9 and GTIs longer than 800 seconds. Therefore, in Table \ref{table2} we present detailed information for all QPO signals detected in each observation, including the energy band, exposure time, corresponding orbital phase, luminosity, and properties of 5 mHz and 10 mHz QPOs.

\par
Utilizing the advantage of wavelet analysis to provide detailed time-frequency information, we found that the transient 10 mHz quasi-periodic signals appeared simultaneously in 1-10 keV and 10-30 keV based on the example ObsID (P010130800507) in Figure \ref{fig:4}. The wavelet results show some detailed structure of QPO signals in different energy bands for both time and frequency domains. In the early stage, both LE and ME data has a short time scale of an oscillation signal near 40 mHz lasting about 100 s. After this 40 mHz signal disappears, the 10 mHz QPOs appear in both energy ranges with a different behavior. The QPO centroid frequencies in the ME (10-30 keV) are generally higher than that in the LE (1-10 keV). Besides, the mHz QPO of LE maintains a steady frequency around 10 mHz for ObsID P010130800507, while the QPO frequency undergoes significant evolution in ME, it starts around 15–18 mHz in the early stages, gradually decreases to near 10 mHz, and eventually splits into two frequencies around 10 mHz and 20 mHz. The QPOs in both energy bands last about 400 s. In Figure \ref{fig:4}, the count rates in LE and ME bands, and the hardness ratio do not show the strong evolution, suggesting that the QPO feature may not be related to luminosity and spectral shapes. A similar energy dependence of QPO frequency is also observed in ObsID P010130800101: the LE data exhibit a transient QPO at 9 mHz for GTIs shorter than 700 seconds, and ME data simultaneously exhibit a transient QPO at 0.014 Hz with the relatively low R factor ($<0.9$, thus not present in Table 2).

\par
\begin{figure}
    \centering
    \subfigure{
        \includegraphics[width=0.46\textwidth, keepaspectratio]{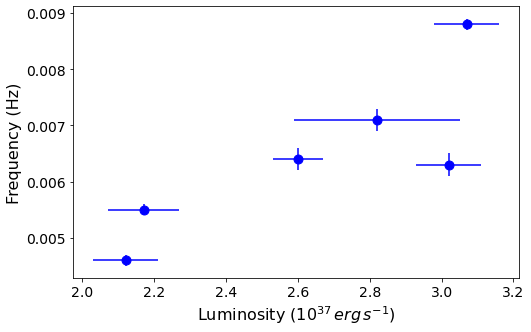}
    }
    \hspace{0.01\textwidth} 
    \subfigure{
        \includegraphics[width=0.46\textwidth, keepaspectratio]{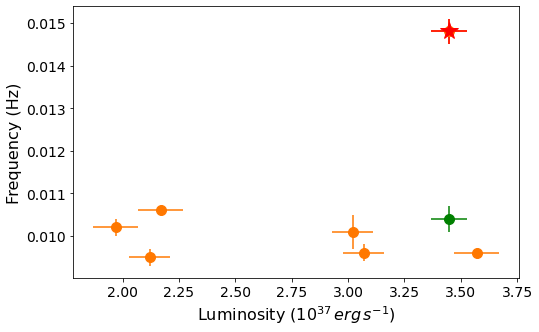}
    }
    
\caption{The QPO frequencies observed in the ME (10-30 keV) versus the 2-110 keV X-ray luminosity of Her X-1. To illustrate the different dependencies on luminosity, $\sim 5$ mHz QPOs in the ME (10-30 keV) are shown as blue dots in the top panel, and $\sim 10$ mHz QPOs are shown as orange dots in the bottom panel. The observation marked with a red star is the average centroid frequency of the PDS for ObsID 0507 which also exhibits a significant evolution in QPO frequency from wavelet analysis (in the range of 0.010-0.018 Hz for the ME band, see Figure \ref{fig:4}), with the FWHM in the PDS reaching approximately $\sim 4.5$ mHz, nearly twice the average value observed in other observations. For a comparison, the green dots indicate the stable QPO frequency near 10 mHz detected in the 1-10 keV band for ObsID 0507 (see Table \ref{table2}).
 }
    \label{fig:5}
\end{figure}

Based on the simple energy spectral fitting with the model of $TBabs\times cflux\times cutoffpl$ for the available observations, we derive the X-ray luminosity from 2-110 keV (assuming the distance of 6.6 kpc) for each observation, thus we also plot the relationships between luminosity and QPO frequencies derived from ME and LE  data in Figure \ref{fig:5}. The 5 mHz QPOs exhibit a positive correlation with X-ray luminosity, with the centroid frequency shifting from 5 mHz to 9 mHz as the luminosity increases from $\sim 2\times 10^{37}erg\,\,s^{-1}$ to $\sim 4 \times 10^{37}erg\,\,s^{-1}$. The $\sim$10 mHz QPOs exhibit a nearly stable frequency over luminosity, except for the ObsID 0507 (also see Table \ref{table2}). The ME data in the ObsID 0507 show an average QPO frequency near 0.015 Hz, with the FWHM reaching approximately $\sim$ 4.5 mHz, nearly twice the average value observed in other observations. In addition, the wavelet analysis suggests that the QPO frequency shows a strong evolution from $\sim 0.018$ to 0.010 Hz in ME band, while the QPO frequency in LE band keeps constant at $\sim 0.01$ Hz. Therefore, the 10 mHz QPO frequency would be not dependent on luminosity evolution. Since two types of QPOs at $\sim 5$ and 10 mHz exhibit distinct dependencies on luminosity and approximately equal global wavelet power (see details in Table \ref{table2}), we infer that different physical mechanisms are responsible for the two types of mHz QPOs.

\begin{table*}
\centering
\caption{Summary of global parameters of QPOs observed in Her X-1 with Insight-HXMT from 2017 to 2019. Q: quality factor, R: relative to the global 95$\%$ confidence spectrum. The observation ID(obsID) are the last four digits of P01013080.}
\label{table2}
\renewcommand{\arraystretch}{1.5}
\fontsize{12pt}{15pt}\selectfont
\resizebox{\textwidth}{!}{ 
\begin{tabular}{c|cccccccccccc}
\hline
ObsID & \begin{tabular}{c}Energy \\ (keV)\end{tabular} & \begin{tabular}{c}Start\\ (MJD)\end{tabular} & \begin{tabular}[c]{@{}c@{}}Exposure\\ (s)\end{tabular} & \begin{tabular}[c]{@{}c@{}}Orbital \\ Phase\end{tabular} & \begin{tabular}[c]{@{}c@{}}35-day \\ phase\end{tabular} & \begin{tabular}{c}Luminosity(2-110 keV) \\($10^{37} erg$ s$^{-1}$)\end{tabular} & \begin{tabular}[c]{@{}c@{}}QPO centroid \\ Frequency\\ (Hz)\end{tabular} & Q factor & R factor & \begin{tabular}[c]{@{}c@{}}QPO centroid\\ Frequency\\ (Hz)\end{tabular} & Q factor & R factor \\ \hline
0207 & 10-30 & 57971.10 & 1860 & 0.46 & 0.03 & 2.12(9) & 0.0046(1) & 4.18(9) & 0.97(7) & 0.0095(2) & 4.13(17) & 0.95(8) \\ \hline
0402 & 10-30 & 58139.11 & 1080 & 0.28 & 0.85 & 3.02(9) & 0.0063(2) & 3.71(24) & 0.92(11) & 0.0101(4) & 4.04(8) & 1.22(4) \\ \hline
0406 & 10-30 & 58139.63 & 840 & 0.59 & 0.86 & 3.07(9) & - & - & - & 0.0096(2) & 4.17(13) & 1.08(8) \\ \hline
0504 & 10-30 & 58140.69 & 840 & 0.21 & 0.90 & 3.57(10) & - & - & - & 0.0096(1) & 5.05(5) & 1.22(3) \\ \hline
\multirow{2}{*}{0507} & 1-10 & 58141.09 & 890 & 0.45 & 0.91 & \multirow{2}{*}{3.45(8)} & - & - & - & 0.0104(3) & 4.00(16) & 1.02(9) \\
 & 10-30 & 58141.09 & 990 & 0.45 & 0.91 &  & - & - & - & 0.0148(4) & 3.36(14) & 1.07(8) \\ \hline
0702 & 10-30 & 58243.21 & 1140 & 0.51 & 0.84 & 3.07(9) & 0.0088(1) & 4.19(9) & 1.43(5) & - & - & - \\ \hline
0801 & 10-30 & 58244.73 & 1109 & 0.40 & 0.88 & 2.17(10) & 0.0055(1) & 5.00(18) & 1.15(8) & 0.0106(2) & 3.03(11) & 1.00(6) \\ \hline
0803 & 10-30 & 58244.99 & 1620 & 0.56 & 0.89 & 1.97(10) & - & - & - & 0.0102(1) & 3.45(3) & 0.94(2) \\ \hline
1303 & 10-30 & 58276.54 & 1620 & 0.12 & 0.79 & 2.60(7) & 0.0064(2) & 2.67(12) & 1.01(6) & - & - & - \\ \hline
1504 & 10-30 & 58278.66 & 1770 & 0.36 & 0.86 & 2.82(23) & 0.0071(2) & 2.15(12) & 1.09(6) & - & - & - \\ \hline
\end{tabular}}
\end{table*}

\section{DISCUSSION}
\label{DISCUSSION}
The mHz QPOs have been commonly observed in both transient and persistent populations of accretion-powered X-ray pulsars \citep{paul2011transient}. This intermediate-mass system Her X-1 shows similar behavior to HMXBs: the presence of highly magnetized neutron stars causes the inner disk radius to extend to approximately 1000 km, leading to Keplerian frequencies in the mHz range. As a result, we can anticipate detecting these X-ray mHz QPOs (approximately 1 mHz–1 Hz, \citealt{psaltis2004accreting}) in Her X-1. In the previous section, we have identified mHz X-ray quasiperiodic oscillations around $\sim 5$ and 10 mHz and periodic pulsations from Her X-1 in both the power density spectrum and wavelet power spectrum. The discovery of X-ray mHz QPOs together with the previous report of UV QPOs in the same source, should be important to understand the physical origin of mHz QPOs in X-ray pulsars. This is unlike mHz QPOs observed in LMXBs, which are connected with some sort of nuclear burnings on the neutron star surface \citep{lyu2015spectral}. Here, we first aim to present a possible theoretical explanation for the mHz QPOs for Her X-1.
\par
Various models have been proposed to explain the characteristics of mHz QPOs in HMXBs, among which the most widely accepted scenario attributes these oscillations to plasma instabilities generated around the magnetospheric boundary. Two notable models are the Keplerian Frequency Model (KFM; \citep{van1987intensity} ) and the Beat Frequency Model (BFM; \citep{alpar1985gx5}). In the KFM, QPOs are generated by inhomogeneities in the inner accretion disc at the Keplerian frequency. Thus, the Keplerian frequency is $\nu_k$ = $\nu_{qpo}$. However, the KFM is only applicable when the neutron star's spin frequency is lower than the Keplerian frequency at the radius at which the QPO-generating inhomogeneity is located due to a faster rotating NS will give rise to centrifugal forces that will inhibit accretion. This scenario explains the case for some of the HMXBs like EXO 2030+375 \citep{angelini1989discovery}. The QPO frequency due to BFM is the difference between the spin frequency and the Keplerian frequency of the inner edge of the accretion disc, meaning that $\nu_{qpo}$ = $\nu_{k}$-$\nu_{spin}$.  Additionally, the BFM and KFM is used to explain the mHz QPOs observed for several pulsars \citep{manikantan2024energy}, such as IGR J19294+1816 (30 mHz in 0.5-10 keV), V 0332+53 (40 mHz in 0.5-60 keV), XTE J1858 + 034 (180 mHz in 3-25 keV).
\par
Her X-1 has a spin frequency of $\nu_{spin}$= 0.8 Hz, and the measured QPO frequency lies at $\nu_{qpo}$ $\sim$ 0.005-0.015 Hz. Since the $\nu _{spin}>\nu _{qpo}$, the KFM does not apply to this object in this scenario. To examine if the BFM is applicable, which assumes the generation of QPOs at the $Alf\acute{e}n$ radius $R_{m}$ \citep{ghosh1979accretion}, we obtain a magnetic field strength of $~2.9\times 10^{12}\,\,G$ with the mass and radius of $10^6$ cm and $1.4M_{\odot}$, respectively. The $Alf\acute{e}n$ radius is about $3.5\times 10^7cm$ to $4.1\times 10^8cm$ which is also in agreement with the radius of $1.9\times 10^8$ cm estimated for QPO generation based on the BFM.

Subsequently, we attempt to interpret the $\sim 10$ mHz QPO in Her X-1 as a beat frequency. Due to higher-frequency QPOs being associated with a smaller radius in the accretion flow, an increase in the QPO frequency with higher energies could be expected. Therefore, the $\sim 10$ mHz QPO observed in the 10–30 keV appears to have a higher frequency in the 1–10 keV occurring in relatively outer regions as seen in Figure \ref{fig:4}. Meanwhile, the $Alfv\acute{e}n$ radius of Her X-1 must be close to the corotation radius, causing the instability of the accretion flow near the boundary with variable inner radius relative to the pulsar, which would lead to fast variations of the $\sim 10$ mHz QPO in the 10–30 keV. 

The 10 mHz X-ray QPOs are also comparable to the previously detected UV QPOs including the centroid frequency, rms amplitude, and Q factor. As noted by \citet{chakrabarty1998high}, the identification of the optical/UV QPOs at the same frequency in X-ray band suggests that a considerable fraction of the optical/UV luminosity from the binary system is a direct reflection of the X-ray emission and the optical/UV flux variations most likely originate as a reprocessing of the X-ray emission on the surface of the mass donor and the accretion disk \citep{middleditch19814u}. Therefore, the QPOs detected in both the X-ray and optical/UV bands at the similar frequency may provide a powerful tool for probing geometry of the binary system and physics of the accretion disk.
\par
Two QPOs at $\sim5$ and $\sim10$ mHz show different behaviors in short time scales based on wavelet results, and have approximately equal global wavelet power (refer to Table \ref{table2} for details), then it would be problematic to just attribute the $\sim10$ mHz QPOs to the harmonics of $\sim5$ mHz QPOs. In addition, the 5 mHz QPO frequency has a correlation with X-ray luminosity, while the 10 mHz QPO has no relation to the luminosity, then the 5 mHz QPOs would have a different physical origin. Another possibility for causing such low-frequency mHz QPOs is the magnetic disk precession model \citep{shirakawa2002magnetically}. In this scenario, the inner region of the accretion disk is subject to magnetic torques that can cause warping and precession of the accretion disk. Under typical conditions in X-ray pulsars, these magnetic torques can overcome the viscous damping and enable the instability mode to grow, which may in turn generate mHz QPOs. The QPO precessional frequency as specified by Equation (27) in \citet{shirakawa2002magnetically} is

\begin{equation}
t_{\mathrm{prec}}\simeq 
775.9\left(\frac{L x}{10^{37} \mathrm{erg} \mathrm{~s}^{-1}}\right)^{-0.71}\left(\frac{\alpha}{0.1}\right)^{0.85} \rm s,
\label{equation11}
\end{equation}
where $\alpha$ is the accretion disk viscosity parameter, and $L_{37}$ is the X-ray luminosity in units of $10^{37}$ erg s$^{-1}$. Assuming an $\alpha$ =0.023 (this choice comes from the model fits carried out by \citet{roy2019laxpc} for the source 4U 0115+63), we obtain a predicted $\nu _{qpo}\sim 6$ mHz at $\sim 2.0\times10^{37}$ erg s$^{-1}$ and the central frequency of the QPO changes to 9 mHz when the source reached an X-ray luminosity of $\sim 3.1\times10^{37}$ erg s$^{-1}$. Thus, a positive correlation between QPO centroid frequency and the X-ray luminosity seen in $\sim5-9$ mHz QPOs may imply that the QPO features are caused by this mechanism.

Previously, the precession of the accretion disk in Her X-1 causing its warped edge to temporarily obscure the pulsar and lead to variations in luminosity (e.g., superorbital period) has been a mainstream hypothesis \citep{brumback2021broadband}. Recent IXPE observations have revealed that the approximately free precession of the modestly asymmetric crust drives the 35-day superorbital period, the spin axis of the crust has shifted relative to the magnetic axis of the neutron star, resulting in a change in the angle between the magnetic axis and the spin axis ($\theta$) within the superorbital period \citep{heyl2024complex}. Magnetic torques that induce warping and precession of the disk are finite and vary with the $\theta$ \citep{lai1999magnetically}. As the crust executes its precessional motion, the torque exerted on the disk by the magnetic field of the neutron star will also vary on the superorbital timescale. Consequently, the precession of the crust excites variations in the dielectric properties of the disk, causing the variation of the precession frequency over the superorbital period. A positive correlation between the QPO centroid frequency and the X-ray luminosity observed in $\sim5-9$ mHz QPOs may provide evidence supporting this mechanism.

\section{Summary}
We have analyzed the timing characteristics of the persistent X-ray binary pulsar Her X-1 by using the observations for Insight-HXMT from 2017 to 2019. The 10 mHz X-ray QPOs that are comparable to the previously detected UV QPOs from the source are detected in the power density spectrum and wavelet power spectrum. We first identified QPOs at $\sim 0.005-0.009$ Hz coexisting with the $\sim$ 0.01 Hz QPOs in the light curve of Her X-1 within the 10–30 keV energy band in the wavelet power spectrum.  The main findings show that the $\sim5-9$ mHz QPO is positively correlated with luminosity, while the 10 mHz QPO has a nearly constant frequency over luminosity. The 10 mHz X-ray QPOs can be interpreted as a beat frequency in the Her X-1, the $Alfv\acute{e}n$ radius should be close to the corotation radius, and 10 mHz UV variability comes from the reprocessing of variable X-ray emission. The 5 mHz QPOs have the different origin from the 10 mHz feature, and would be caused by magnetic disk precession.

\section*{Acknowledgements}
We are grateful to the referee for the suggestions to improve the manuscript. This work is supported by the the NSFC (No. 12133007) and National Key Research and Development Program of China (Grants No. 2021YFA0718503, 2023YFA1607901). This work has made use of data from the Insight-HXMT mission, a project funded by China National Space Administration (CNSA) and the Chinese Academy of Sciences (CAS).
\bibliography{sample631}{}

\begin{thebibliography}{}
\expandafter\ifx\csname natexlab\endcsname\relax\def\natexlab#1{#1}\fi
\providecommand{\url}[1]{\href{#1}{#1}}
\providecommand{\dodoi}[1]{doi:~\href{http://doi.org/#1}{\nolinkurl{#1}}}
\providecommand{\doeprint}[1]{\href{http://ascl.net/#1}{\nolinkurl{http://ascl.net/#1}}}
\providecommand{\doarXiv}[1]{\href{https://arxiv.org/abs/#1}{\nolinkurl{https://arxiv.org/abs/#1}}}

\bibitem[{Addison(2017)}]{addison2017illustrated}
Addison, P.~S. 2017, The illustrated wavelet transform handbook: introductory theory and applications in science, engineering, medicine and finance (CRC press)

\bibitem[{Alpar \& Shaham(1985)}]{alpar1985gx5}
Alpar, M.~A., \& Shaham, J. 1985, Nature, 316, 239

\bibitem[{Angelini {et~al.}(1989)Angelini, Stella, \& Parmar}]{angelini1989discovery}
Angelini, L., Stella, L., \& Parmar, A. 1989, Astrophysical Journal, Part 1 (ISSN 0004-637X), vol. 346, Nov. 15, 1989, p. 906-911., 346, 906

\bibitem[{Boroson {et~al.}(2000)Boroson, O’Brien, Horne, Kallman, Still, Boyd, Quaintrell, \& Vrtilek}]{boroson2000discovery}
Boroson, B., O’Brien, K., Horne, K., {et~al.} 2000, The Astrophysical Journal, 545, 399

\bibitem[{Brumback {et~al.}(2021)Brumback, Hickox, F{\"u}rst, Pottschmidt, Tomsick, Wilms, Staubert, \& Vrtilek}]{brumback2021broadband}
Brumback, M.~C., Hickox, R.~C., F{\"u}rst, F.~S., {et~al.} 2021, The Astrophysical Journal, 909, 186

\bibitem[{Bu {et~al.}(2015)Bu, Chen, Li, Qu, Belloni, \& Zhang}]{bu2015correlations}
Bu, Q.-c., Chen, L., Li, Z.-s., {et~al.} 2015, The Astrophysical Journal, 799, 2

\bibitem[{Cao {et~al.}(2020)Cao, Jiang, Meng, Zhang, Luo, Yang, Zhang, Gu, Sun, Liu, {et~al.}}]{cao2020medium}
Cao, X., Jiang, W., Meng, B., {et~al.} 2020, SCIENCE CHINA Physics, Mechanics \& Astronomy, 63, 1

\bibitem[{Chakrabarty(1998)}]{chakrabarty1998high}
Chakrabarty, D. 1998, The Astrophysical Journal, 492, 342

\bibitem[{Chen {et~al.}(2022)Chen, Wang, You, Tian, Liu, Zhang, Ding, Qu, Zhang, Song, {et~al.}}]{chen2022wavelet}
Chen, X., Wang, W., You, B., {et~al.} 2022, Monthly Notices of the Royal Astronomical Society, 513, 4875

\bibitem[{Chen {et~al.}(2020)Chen, Cui, Li, Wang, Xu, Lu, Wang, Chen, Han, Hu, {et~al.}}]{chen2020low}
Chen, Y., Cui, W., Li, W., {et~al.} 2020, Science China Physics, Mechanics \& Astronomy, 63, 1

\bibitem[{Devasia {et~al.}(2011)Devasia, James, Paul, \& Indulekha}]{devasia2011rxte}
Devasia, J., James, M., Paul, B., \& Indulekha, K. 2011, Monthly Notices of the Royal Astronomical Society, 414, 1023

\bibitem[{Ding {et~al.}(2021)Ding, Wang, Zhang, Bu, Cai, Cao, Zhi, Chen, Chen, Chen, {et~al.}}]{ding2021qpos}
Ding, Y., Wang, W., Zhang, P., {et~al.} 2021, Monthly Notices of the Royal Astronomical Society, 503, 6045

\bibitem[{Finger {et~al.}(1996)Finger, Wilson, \& Harmon}]{finger1996quasi}
Finger, M., Wilson, R., \& Harmon, B. 1996, Astrophysical Journal v. 459, p. 288, 459, 288

\bibitem[{Ghosh {et~al.}(2023)Ghosh, Gallo, \& Gonzalez}]{ghosh2023applying}
Ghosh, A., Gallo, L., \& Gonzalez, A. 2023, Monthly Notices of the Royal Astronomical Society, 524, 1478

\bibitem[{Ghosh \& Lamb(1979)}]{ghosh1979accretion}
Ghosh, P., \& Lamb, F. 1979, Astrophysical Journal, Part 1, vol. 234, Nov. 15, 1979, p. 296-316., 234, 296

\bibitem[{Heyl {et~al.}(2024)Heyl, Doroshenko, Gonz{\'a}lez-Caniulef, Caiazzo, Poutanen, Mushtukov, Tsygankov, Kirmizibayrak, Bachetti, Pavlov, {et~al.}}]{heyl2024complex}
Heyl, J., Doroshenko, V., Gonz{\'a}lez-Caniulef, D., {et~al.} 2024, Nature Astronomy, 1

\bibitem[{Inam {et~al.}(2004)Inam, Baykal, Swank, \& Stark}]{inam2004discovery}
Inam, S., Baykal, A., Swank, J., \& Stark, M. 2004, The Astrophysical Journal, 616, 463

\bibitem[{James {et~al.}(2010)James, Paul, Devasia, \& Indulekha}]{james2010discovery}
James, M., Paul, B., Devasia, J., \& Indulekha, K. 2010, Monthly Notices of the Royal Astronomical Society, 407, 285

\bibitem[{Jimenez-Garate {et~al.}(2002)Jimenez-Garate, Hailey, den Herder, Zane, \& Ramsay}]{jimenez2002high}
Jimenez-Garate, M., Hailey, C., den Herder, J., Zane, S., \& Ramsay, G. 2002, arXiv preprint astro-ph/0206181

\bibitem[{Kaur {et~al.}(2007)Kaur, Paul, Raichur, \& Sagar}]{kaur2007quasi}
Kaur, R., Paul, B., Raichur, H., \& Sagar, R. 2007, The Astrophysical Journal, 660, 1409

\bibitem[{Lai(1999)}]{lai1999magnetically}
Lai, D. 1999, The Astrophysical Journal, 524, 1030

\bibitem[{Leahy(2003)}]{leahy2003modelling}
Leahy, D. 2003, Monthly Notices of the Royal Astronomical Society, 342, 446

\bibitem[{Leahy \& Igna(2011)}]{leahy2011light}
Leahy, D., \& Igna, C. 2011, The Astrophysical Journal, 736, 74

\bibitem[{Leahy {et~al.}(2020)Leahy, Postma, \& Chen}]{leahy2020astrosat}
Leahy, D., Postma, J., \& Chen, Y. 2020, The Astrophysical Journal, 889, 131

\bibitem[{Leahy \& Wang(2020)}]{leahy2020swift}
Leahy, D., \& Wang, Y. 2020, The Astrophysical Journal, 902, 146

\bibitem[{Liu {et~al.}(2020)Liu, Zhang, Li, Lu, Chang, Li, Zhang, Jin, Yu, Zhang, {et~al.}}]{liu2020high}
Liu, C., Zhang, Y., Li, X., {et~al.} 2020, SCIENCE CHINA Physics, Mechanics \& Astronomy, 63, 1

\bibitem[{Liu {et~al.}(2022)Liu, Wang, Chen, Yang, Lu, Song, Qu, Zhang, \& Zhang}]{liu2022detection}
Liu, Q., Wang, W., Chen, X., {et~al.} 2022, Monthly Notices of the Royal Astronomical Society, 516, 5579

\bibitem[{Lyu {et~al.}(2015)Lyu, M{\'e}ndez, Zhang, \& Keek}]{lyu2015spectral}
Lyu, M., M{\'e}ndez, M., Zhang, G., \& Keek, L. 2015, Monthly Notices of the Royal Astronomical Society, 454, 541

\bibitem[{Manikantan {et~al.}(2024)Manikantan, Paul, Sharma, Pradhan, \& Rana}]{manikantan2024energy}
Manikantan, H., Paul, B., Sharma, R., Pradhan, P., \& Rana, V. 2024, Monthly Notices of the Royal Astronomical Society, 531, 530

\bibitem[{Middleditch {et~al.}(1981)Middleditch, Mason, Nelson, \& White}]{middleditch19814u}
Middleditch, J., Mason, K., Nelson, J., \& White, N. 1981, Astrophysical Journal, 244, 1001

\bibitem[{Moon \& Eikenberry(2001)}]{moon2001discovery}
Moon, D.-S., \& Eikenberry, S.~S. 2001, The Astrophysical Journal, 552, L135

\bibitem[{O'Brien {et~al.}(2001)O'Brien, Horne, Boroson, Still, Gomer, Oke, Boyd, \& Vrtilek}]{o2001keck}
O'Brien, K., Horne, K., Boroson, B., {et~al.} 2001, Monthly Notices of the Royal Astronomical Society, 326, 1067

\bibitem[{Paul \& Naik(2011)}]{paul2011transient}
Paul, B., \& Naik, S. 2011, arXiv preprint arXiv:1110.4446

\bibitem[{Paul \& Rao(1998)}]{paul1998quasi}
Paul, B., \& Rao, A. 1998, arXiv preprint astro-ph/9805366

\bibitem[{Petterson(1975)}]{petterson1975hercules}
Petterson, J.~A. 1975, Astrophysical Journal, vol. 201, Oct. 15, 1975, pt. 2, p. L61-L64., 201, L61

\bibitem[{Psaltis(2004)}]{psaltis2004accreting}
Psaltis, D. 2004, arXiv preprint astro-ph/0410536

\bibitem[{Raman {et~al.}(2021)Raman, Paul, \& Bhattacharya}]{raman2021astrosat}
Raman, G., Paul, B., \& Bhattacharya, D. 2021, Monthly Notices of the Royal Astronomical Society, 508, 5578

\bibitem[{Reynolds {et~al.}(1997)Reynolds, Quaintrell, Still, Roche, Chakrabarty, \& Levine}]{reynolds1997new}
Reynolds, A., Quaintrell, H., Still, M., {et~al.} 1997, Monthly Notices of the Royal Astronomical Society, 288, 43

\bibitem[{Roy {et~al.}(2019)Roy, Agrawal, Iyer, Bhattacharya, Yadav, Antia, Chauhan, Choudhury, Dedhia, Katoch, {et~al.}}]{roy2019laxpc}
Roy, J., Agrawal, P., Iyer, N., {et~al.} 2019, The Astrophysical Journal, 872, 33

\bibitem[{Scott \& Leahy(1999)}]{scott1999rossi}
Scott, D., \& Leahy, D. 1999, The Astrophysical Journal, 510, 974

\bibitem[{Shirakawa \& Lai(2002)}]{shirakawa2002magnetically}
Shirakawa, A., \& Lai, D. 2002, The Astrophysical Journal, 565, 1134

\bibitem[{Staubert {et~al.}(1983)Staubert, Bezler, \& Kendziorra}]{staubert1983hercules}
Staubert, R., Bezler, M., \& Kendziorra, E. 1983, Astronomy and Astrophysics, vol. 117, no. 2, Jan. 1983, p. 215-219., 117, 215

\bibitem[{Staubert {et~al.}(2009)Staubert, Klochkov, \& Wilms}]{staubert2009updating}
Staubert, R., Klochkov, D., \& Wilms, J. 2009, arXiv preprint arXiv:0904.2307

\bibitem[{Takeshima {et~al.}(1994)Takeshima, Dotani, Mitsuda, \& Nagase}]{takeshima1994discovery}
Takeshima, T., Dotani, T., Mitsuda, K., \& Nagase, F. 1994, Astrophysical Journal, Part 1 (ISSN 0004-637X), vol. 436, no. 2, p. 871-874, 436, 871

\bibitem[{Tananbaum {et~al.}(1972)Tananbaum, Gursky, Kellogg, Levinson, Schreier, \& Giacconi}]{tananbaum1972discovery}
Tananbaum, H.~a., Gursky, H., Kellogg, E., {et~al.} 1972, Astrophysical Journal, vol. 174, p. L143, 174, L143

\bibitem[{Torrence \& Compo(1998)}]{torrence1998practical}
Torrence, C., \& Compo, G.~P. 1998, Bulletin of the American Meteorological society, 79, 61

\bibitem[{Tr{\"u}mper {et~al.}(1978)Tr{\"u}mper, Pietsch, Reppin, Voges, Staubert, \& Kendziorra}]{trumper1978evidence}
Tr{\"u}mper, J., Pietsch, W., Reppin, C., {et~al.} 1978, Astrophysical Journal, Part 2-Letters to the Editor, vol. 219, Feb. 1, 1978, p. L105-L110. Deutsche Forschungsgemeinschaft, 219, L105

\bibitem[{Van~der Klis {et~al.}(1987)Van~der Klis, Stella, White, Jansen, \& Parmar}]{van1987intensity}
Van~der Klis, M., Stella, L., White, N., Jansen, F., \& Parmar, A. 1987, Astrophysical Journal, Part 1 (ISSN 0004-637X), vol. 316, May 1, 1987, p. 411-426., 316, 411

\bibitem[{Vrtilek {et~al.}(2001)Vrtilek, Quaintrell, Boroson, Still, Fiedler, O’Brien, \& McCray}]{vrtilek2001multiwavelength}
Vrtilek, S., Quaintrell, H., Boroson, B., {et~al.} 2001, The Astrophysical Journal, 549, 522

\bibitem[{Wang {et~al.}(2021)Wang, Tang, Tuo, Epili, Zhang, Song, Lu, Qu, Zhang, Ge, Huang, Li, Bu, Cai, Cao, Chang, Chen, Chen, Chen, Chen, Chen, Cui, Du, Gao, Gao, Gu, Guan, Guo, Han, Huo, Jia, Jiang, Jin, Kong, Li, Li, Li, Li, Li, Li, Li, Li, Liang, Liao, Liu, Liu, Liu, Liu, Lu, Luo, Luo, Ma, Ma, Meng, Nang, Nie, Ou, Ren, Sai, Song, Sun, Tao, Wang, Wang, Wang, Wang, Wang, Wen, Wu, Wu, Wu, Xiao, Xiao, Xiong, Xu, Yang, Yang, Yang, Yang, Yi, Yin, You, Zhang, Zhang, Zhang, Zhang, Zhang, Zhang, Zhang, Zhang, Zhao, Zhao, Zheng, Zheng, \& Zhou}]{WANG20211}
Wang, W., Tang, Y., Tuo, Y., {et~al.} 2021, Journal of High Energy Astrophysics, 30, 1, \dodoi{https://doi.org/10.1016/j.jheap.2021.01.002}

\bibitem[{Xiao {et~al.}(2019)Xiao, Ji, Staubert, Ge, Zhang, Zhang, Santangelo, Ducci, Liao, Guo, {et~al.}}]{xiao2019constant}
Xiao, G., Ji, L., Staubert, R., {et~al.} 2019, Journal of High Energy Astrophysics, 23, 29

\bibitem[{Zhang {et~al.}(2014)Zhang, Lu, Zhang, \& Li}]{zhang2014introduction}
Zhang, S., Lu, F., Zhang, S., \& Li, T. 2014, in Space Telescopes and Instrumentation 2014: Ultraviolet to Gamma Ray, Vol. 9144, SPIE, 588--595

\bibitem[{Zhu \& Wang(2024)}]{zhu2024energy}
Zhu, H., \& Wang, W. 2024, The Astrophysical Journal, 968, 106

\end{thebibliography}
\bibliographystyle{aasjournal}
\end{document}